\documentclass[preprint]{aastex63}

\usepackage{amsmath}
\usepackage{amssymb}
\usepackage{latexsym}
\usepackage{graphicx}
\usepackage{color}

\begin{document}
\title{Transverse motions in sunspot super-penumbral fibrils}

\author{R. J. Morton}
\affiliation{Department of Mathematics, Physics and Electrical Engineering, Northumbria University, Newcastle Upon Tyne,NE1 8ST, UK}

\author{K. Mooroogen}
\affiliation{Department of Mathematics, Physics and Electrical Engineering, Northumbria University, Newcastle Upon Tyne,NE1 8ST, UK}

\author{V. M. J. Henriques}
\affiliation{Institute of Theoretical Astrophysics, University of Oslo, P.O. Box 1029 Blindern, N-0315 Oslo, Norway}
\affiliation{Rosseland Centre for Solar Physics, University of Oslo, P.O. Box 1029 Blindern, N-0315 Oslo, Norway}
\affiliation{Astrophysics Research Centre (ARC), School of Mathematics and Physics, Queen's University Belfast, BT7 1NN, Belfast, Northern Ireland, UK}
%\subject{a}

% \corres{R. J. Morton\\
% \email{richard.morton@northumbria.ac.uk}}

 \begin{abstract}
Sunspots have played a key role in aiding our understanding of magnetohydrodynamic (MHD) wave phenomenon in the Sun's atmosphere, and 
it is well known they demonstrate a number of wave phenomenon associated with slow MHD modes. 
Recent studies have shown that transverse wave modes are present throughout the majority of the chromosphere. 
Using high-resolution Ca II 8542~{\AA} observations from the Swedish Solar Telescope, we provide the first demonstration that {the chromospheric super-penumbral fibrils, which span out from the sunspot, also show ubiquitous transverse motions. We interpret these motions as transverse waves, in particular the MHD kink mode.}
We compile the statistical properties of over 2000 transverse motions to find distributions for periods and amplitudes, finding they are broadly consistent with previous observations of chromospheric transverse waves in quiet Sun fibrils. The very presence of the waves in super-penumbral fibrils raises important questions about how they are generated, and could have implications for our understanding
of how MHD wave energy is transferred through the atmosphere of a sunspot.
 \end{abstract}

\date{Received /Accepted}

%\shorttitle{Waves in sunspots}
%\shortauthors{Morton et al.}

% \maketitle

\section{Introduction}
It is well established that magnetohydrodynamic (MHD) waves are ubiquitous throughout the Sun's atmosphere, with evidence that
they play key roles in energy transfer. A clear and beautiful example of their defining impact is found in the visual appearance and temporal evolution of the chromosphere, with slow MHD shocks
driving chromospheric jets and mass 
loading events \citep{DEPetal2004}. More recently, the identification of Alfv\'enic modes in the chromosphere and corona suggest these modes could supply a significant fraction of the required energy to explain the heating of the quiescent corona and acceleration of the solar wind
\citep{DEPetal2007, TOMetal2007, JESetal2009, MCIetal2011, MORetal2012, MORetal2015, MORetal2019}.

\medskip
One of the outstanding questions around MHD wave propagation throughout the Sun's atmosphere is, how are transverse wave modes excited? The
fluctuations of velocity and magnetic field observed with both
remote and in-situ sensing instruments suggest that Alfv\'enic waves exist over a wide range of time-scales. 
In the solar wind, Alfv\'enic fluctuations have time-scales on the order of at least a few hours \citep[e.g.,][]{Goldstein_1995}. While in the solar chromosphere and corona,
transverse wave modes with Alfv\'enic\footnote{The phrase Alfv\'enic is used to refer to wave modes that have characteristics similar to the pure Alfv\'en mode. To quote \cite{GOOetal2012}, \emph{'[The kink waves are] to a high degree of accuracy incompressible waves with negligible pressure perturbations and with mainly horizontal motions. The main restoring force of kink waves in the magnetised flux tube is the magnetic tension force'}. Hence, kink modes are Alfv\'enic in character.} characteristics (such as the long wavelength fast kink mode\footnote{\cite{GOOetal2012} demonstrate that total pressure perturbations for the kink mode are on the order of $(k_zR)^2$, where $R$ is the flux tube radius and $k_z$ the wavenumber.}) have, to date, only been identified unambiguously with much more 
modest time-scales \citep[$\lesssim 2000$~s, e.g.,][] {OKADEP2011, MORetal2012c, Weberg_2020}.

The current paradigm for the excitation of transverse waves largely focuses on the influence of 
the horizontal component of convection on photospheric magnetic fields. {This mechanism for excitation is
thought only to be relevant for wave excitation in small-scale magnetic elements, i.e., magnetic bright points, that exist with the internetwork and network, as well as plage.} The convection generates transverse
waves with time-scales on the order of $1-10$~minutes, via the buffeting of the magnetic elements \citep{CRAVAN2005}. This pathway for wave energy is a typical assumption for 
many wave-based numerical investigations that examine coronal heating or wind acceleration. However, there remain a number of issues associated with relying solely on transverse wave excitation at the photosphere, not least the large expected rate of reflection at the transition region due to steep gradients in the Alfv\'en speed \citep[for a broader discussion see][]{CAL2017}. 
Moreover, it is not evident how the near-surface convection leads to significant energy input at the larger time-scales 
found in the solar wind. Although, recently \citet{Cranmer_2018} discussed how magnetic reconnection could provide a  contribution to wave generation in the corona from the evolution of the magnetic carpet.

An alternative suggestion that has been receiving increased attention focuses on the role of $p$-modes, which can inject additional energy into the corona through mode conversion. The theoretical description of this process implies that Alfv\'enic modes can be excited at the transition region
\citep{CAL2011, CALHAN2011, KHOCAL2012, CAL2017}. There is also a clear power enhancement found at $\sim4$~mHz in coronal Alfv\'enic velocity fluctuations \citep{MORetal2019} 
that could support the role of $p$-modes in exciting transverse waves. However, further investigation, both observationally and theoretically, is still required to demonstrate the efficacy of this energy pathway. {For sunspots at least, this is potentially the only mechanism that could excite Alfv\'enic waves in the atmosphere that lies above.}

\medskip
Perhaps one of the best features on the Sun for studying MHD
wave phenomenon are sunspots. Through observation and modelling, many aspects of MHD waves in inhomogeneous plasmas have been discovered and understood \citep[e.g.,][]{BOGJUD2006, JESetal2015, JESSetal2020}. The source of MHD
waves found in sunspots is largely considered to be from the internal $p$-modes, leaking out along the magnetic field. The $p$-modes are fast acoustic modes that convert to slow modes across the equipartition layer (where $c_s \approx v_A$). The
slow modes observed in sunspots posses a clear signal of their origin, with a broad frequency
distribution centred around a peak power of 3~mHz \citep[e.g.,][]{Thomas_1984}. The sunspot's strong magnetic fields provide a natural guide for slow magnetoacoustic modes. And their relatively simple structure makes it somewhat straightforward to undertake analysis of observed fluctuations, reducing the complication of disentangling different waves modes (at least in the umbra). 

The slow waves are observed to be channelled into the
chromosphere and steepening as they propagate upwards \citep[e.g.,][]{Lites_1985,Brynildsen_1999}, with the dominant frequency shifting to 5~mHz. Those waves propagating along the more vertical magnetic field in the sunspot umbra, with some wave-fronts steepening to shocks and producing umbral flashes in the chromospheric umbral atmosphere 
\citep{ROUPPEetal2003, TIANetal2014}. In the transition from umbra to penumbra, the dominant oscillatory pattern is the running penumbral waves \citep{Giovanelli_1972}, showing apparent 
propagation outward from the umbra. However, this apparent propagation is largely considered to 
be an illusion arising from upwardly propagating slow waves along inclined field lines \citep{BOGJUD2006, BLOetal2007}. The running penumbral waves show a frequency dependence with 
distance from the umbra, with the largest frequencies in the
inner penumbra and decreasing outwards \citep{BRIZIR1997,KOBMAR2004,Jess_2013}.

Both the shift in frequency of umbral oscillations and the change in dominant frequency of the running penumbral waves can be associated with the influence of gravitational stratification. The stratification of the atmosphere leads to
the presence of an acoustic cut-off, which is modified by the plasma-$\beta$ and the inclination of the magnetic field. 
The expression for the acoustic cut-off frequency, $\nu_{c}$, in a high-$\beta$ plasma is given by \citep{BELLER1977}:

\begin{equation}
    \nu_{c} = \frac{\gamma g\cos\theta_B}{4\pi c_s},
\end{equation}
where $\gamma$ is the ratio of specific heats, $g$ is the gravitational constant, and $c_s$ is the sound speed. Hence,
as the sunspot magnetic field becomes more inclined in the penumbra, this reduces the effective stratification and hence lowers the
cut-off frequency. Therefore lower frequency waves are able to propagate into the chromosphere further from the umbra. 

\medskip
Slow magnetoacoustic modes are not the only MHD mode
expected to be present in the sunspot atmosphere. The inclination of the field enables the conversion of the fast acoustic modes to fast magnetoacoustic modes at the equipartition layer, with the amount of conversion dependent upon inclination and attack angle of the waves \citep[e.g.,][]{SCHCAL2006}. The presence of fast wave modes is now a well established feature of numerical simulations of sunspots \citep[e.g.,][]{KHOCOL2006}, but their presence has so far been inconspicuous in observations. To date, the only signature of their existence has been inferred from an 
enhancement of high-frequency (5.5-7.5~mHz) wave power surrounding active regions \citep{BROetal1992}, which has 
been explained in terms of fast mode refraction due to the increase in Alfv\'en speed in the sunspot atmosphere \citep{Khomenko_2009}. Moreover, the production of fast modes at the equipartition layer is a key step in the generation of Alfv\'enic modes at the transition region.

\medskip

It is hopefully apparent that identifying fast modes in the sunspot atmosphere will be a key step in confirming or challenging the current paradigm around wave energy transfer through $p$-modes. Such results would enable further studies of the typical properties of the fast modes, permitting an assessment of whether existing intuition about mode conversion within a sunspot is correct.
While the fast mode has been difficult to detect through analysis of intensity variations and Doppler velocities,
an examination of the fine-scale structure of the sunspots
chromosphere, i.e., super-penumbral fibrils, may be able to
provide some direct evidence for their existence. 
{Here, we demonstrate that super-penumbral fibrils display ubiquitous transverse motions, which we interpret as the MHD kink mode. {The observed motions are localised to individual fibrils, rather than being a global motion of the sunspot itself}. The presence of kink modes in a structured sunspot atmosphere would provide support for current picture of energy transfer through the mode-conversion of \textit{p}-modes.}

%#############################################################################
%#############################################################################
\begin{figure*}[!tp]
\centering
\includegraphics[scale=0.8, clip=true, viewport=2.cm 6.0cm 19.cm 22.cm]{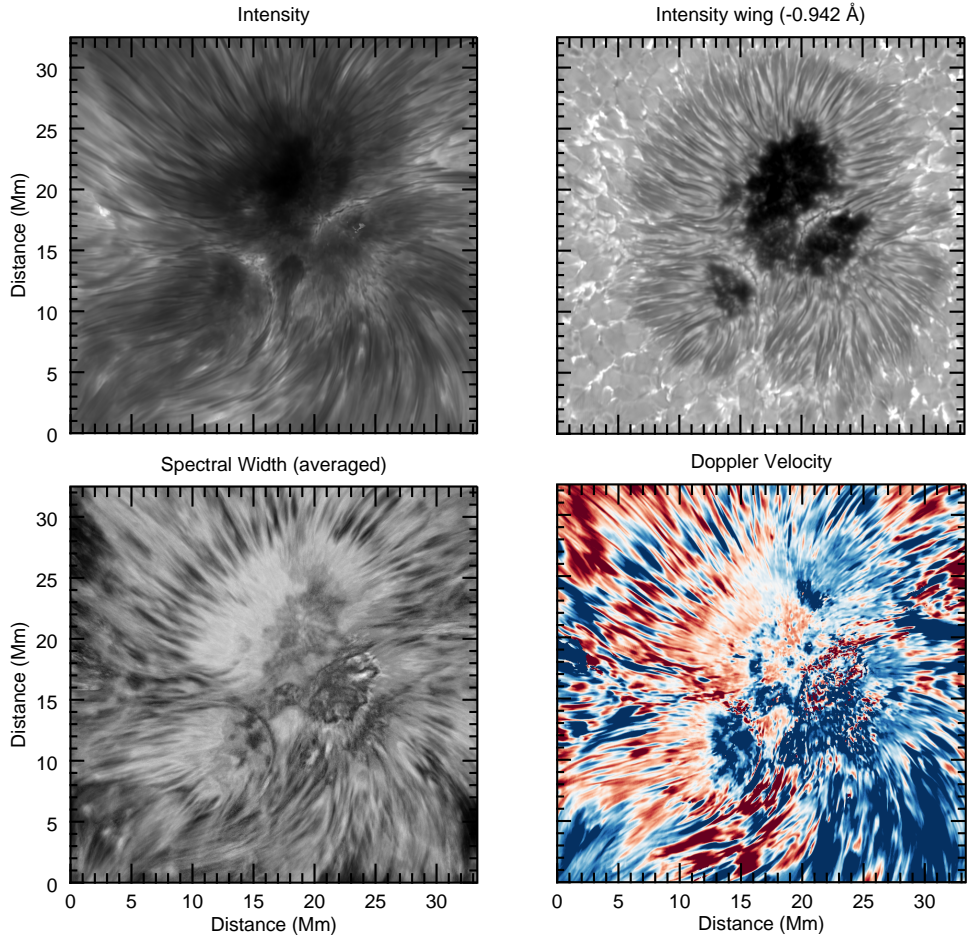}
\caption{The sunspot as observed in Ca II 8542~{\AA}. The upper left image shows the chromospheric emission at the 
nominal line centre wavelength. Super-penumbral fibrils are seen to extended near near-radially from the chromospheric umbra in 
the upper half, while those in the lower half are more curvilinear. The upper right-hand image shows the photospheric section of the sunspot as observed in the wings of the Ca II line (-0.942~{\AA}). The lower two panels show the temporally averaged line width (left) and Doppler velocity (right) data products. The width and Doppler components of the fibrils can be seen in both data product images.}\label{fig:sst_fov}
\end{figure*}
%#############################################################################
%#############################################################################

\section{Observations and data reduction}
The observation presented here are focused on an active region sunspot close to disk centre 
$N~9.25, W~4.50$ (Figure~\ref{fig:sst_fov}), taken by the Swedish Solar Telescope \citep[SST - ]{SCHetal2003} using the 
CRisp Imaging SpectroPolarimeter \citep[CRISP - ][]{SCHA2006, SCHetal2008} on 28 July 2014 
between 10:43 and 11:24~UT. The data set is a full Stokes
polarimetric scan of
the Calcium II 8542~{\AA} infrared triplet line at fifteen wavelength positions in the range  $\pm$(
0.942, 0.580, 0.398, 0.290, 0.217, 0.145, 0.073, 0)~{\AA} from line centre. Further technical details with regards to acquisition and processing can be found in \citet{Henriques_2017}. 

The data has the a spatial sampling  
on the order of $\sim0''.059$ per pixel (spatial resolution $0''.21$). 
The data were processed with an early version of the CRISPRED pipeline \citep{DELetal2015}, centred around multi-object
multi-frame blind deconvolution \citep[MOMFBD][]{VANNetal2005}
using an extended scheme as in \citet{HEN2013}. The data was then de-rotated, aligned, and de-stretched 
\citep{SHIetal1994}.  
 
Here we concentrate on the Stokes I component of the data, and use it to estimate the line profile minimum (LPM) intensity\footnote{The line profile minimum images enable us to disambiguate the velocity effects that are introduced by using the intensity recorded at a single wavelength position in the line scan, e.g., nominal line core.}, Doppler shift and width of the Ca II line (Figure~\ref{fig:sst_fov}). Due to the steep wings of the Ca II line profile, a Voigt profile is used to model the spectral profile for each pixel and all time-frames. The Voigt profile has the form,
 $$
V(\lambda:\sigma, \gamma) = \int^\infty_{-\infty} G(\lambda,\sigma)L(\lambda'-\lambda,\gamma_L)\,d\lambda,
$$
where $G(\lambda, \sigma)$ is a Gaussian profile, 
$L(\lambda'-\lambda, \gamma_L)$ is a Lorentzian and $\lambda'$ is the displacement from line centre. As it can be seen the Voigt profile is a convolution of Gaussian and Lorentzian profiles, taking into account both thermal and collisional broadening mechanisms respectively. These manifest in the parameters, $\gamma_L$ and 
$\sigma$, that refer to the collisional and Doppler broadening mechanisms respectively. Both of these parameters contribute to the width of the line. {The Voigt profile is fit to each Stokes I profile using least-squares minimisation} 
(\verb|mpfit|; - \citealt{MAR2009}). {After fitting the Voigt profile, we use the $\sigma$ parameter as a proxy for exploring thermal variability, whilst ignoring the collision contributions}. There were two missing time-steps in the Stokes I data and there also was a small variation in cadence. 
For each derived data product, the data sequence is homogenised to a cadence of 30~s and missing time-steps filled using spline interpolation for a total of 81 frames.

%\section{Waves in the Sunspot}
\section{Compressive Waves}
As discussed in the introduction, sunspots are well known to host a variety of wave phenomenon, which are often related to the slow magnetoacoustic wave 
modes and thought to be due to the leakage of the internal acoustic modes into the atmosphere. This sunspot is no different. While not our main focus, we give a brief overview of the compressive wave phenomenon that is visible in the
data. Furthermore, we exploit their presence to undertake some basic magneto-seismology that aids our later discussion.

 %#############################################################################
%#############################################################################
\begin{figure*}[!tp]
\centering
\includegraphics[scale=0.8, clip=true, viewport=1.cm 6.5cm 21.cm 21.cm]{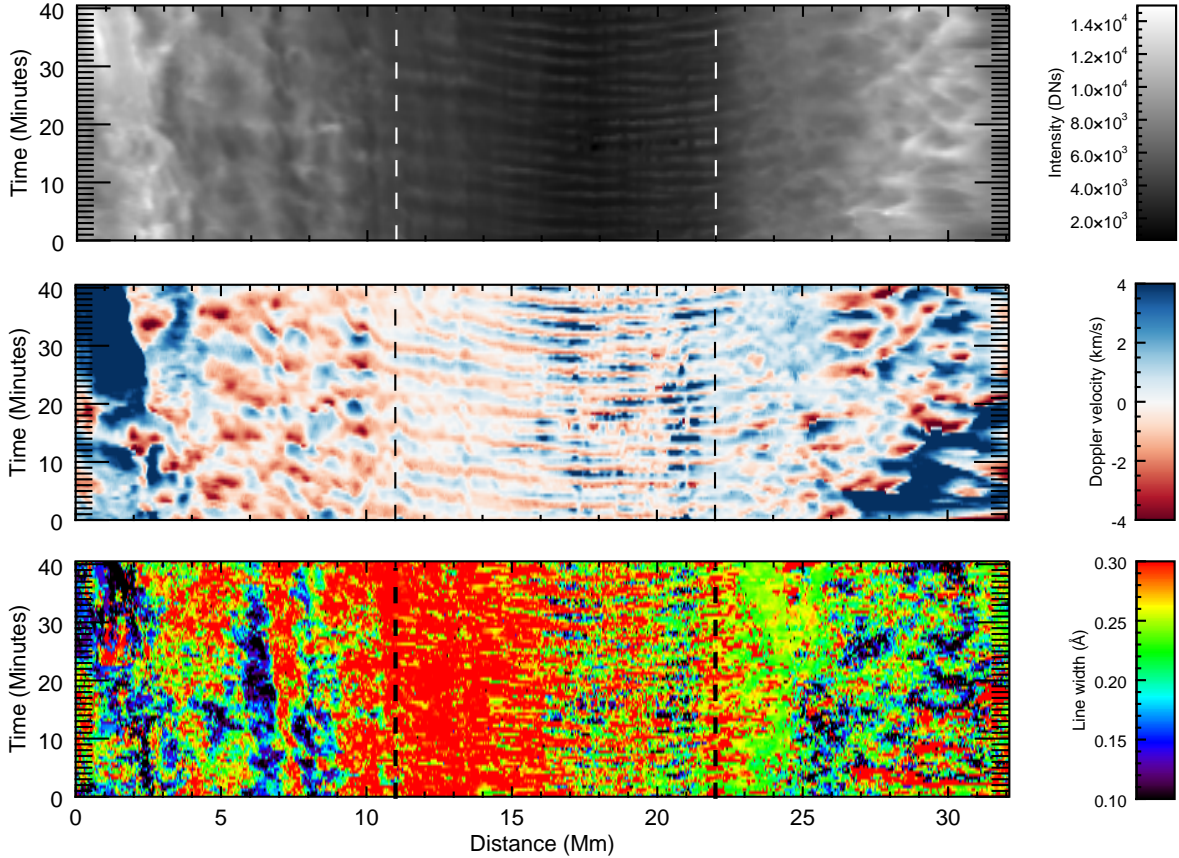}
\caption{Co-spatial time-distance diagrams that reveal the presence of slow magnetoacoustic oscillations. The figure displays trans-sunspot time-distance diagrams of the LPM intensity, Doppler velocity and Doppler line width (from top to bottom respectively). The slow modes can most clearly be identified by the alternating ridged pattern most evident near and through the umbra (area between vertical dotted lines).}\label{fig:slow_waves}
\end{figure*}
%#############################################################################
%#############################################################################

\subsection{Observational signatures}
Figure~\ref{fig:slow_waves} captures the key features on the compressive phenomenon within the umbra and penumbra. In the 
chromospheric umbra we see clear evidence for the upward propagating slow waves, which are marked most strikingly by the periodic red and blue shifts with periods of 6-7~mHz (determined from Fourier analysis of the time-series). There is a 
large-scale bowl-shaped pattern for each wave front in both intensity and Doppler shift (Figure~\ref{fig:slow_waves}). The 
slow waves first appear close to the sunspot centre and show an apparent outward propagation into the penumbral regions. 
The `sides' of the bowl represent the innermost component of the running penumbral waves. As already discussed, this apparent propagation is considered to 
be an illusion arising from upwardly propagating slow waves along inclined field lines \citep{BOGJUD2006,BLOetal2007}. 
The near-vertically propagating waves reaching the region of the atmosphere that contributes to Ca II line formation first, followed by waves propagating along the inclined fields. 

In addition to the spot-wide intensity and velocity fluctuations, there is also visible signatures of the umbral flash phenomenon, marked by saturated blue shifts to the right-hand side of the umbra. We also see that the shocks broaden the Calcium line profile. Due to the mass of the calcium atom, the line is sensitive to non-thermal broadening \citep{CAUZZIetal2009}, hence the observed broadening could either be a signature of plasma heating or turbulence generated by the shocks.

%#############################################################################
\begin{figure*}[!tp]
\centering
\includegraphics[scale=0.6, clip=true, viewport=0.0cm 0cm 24.5cm 8.5cm]{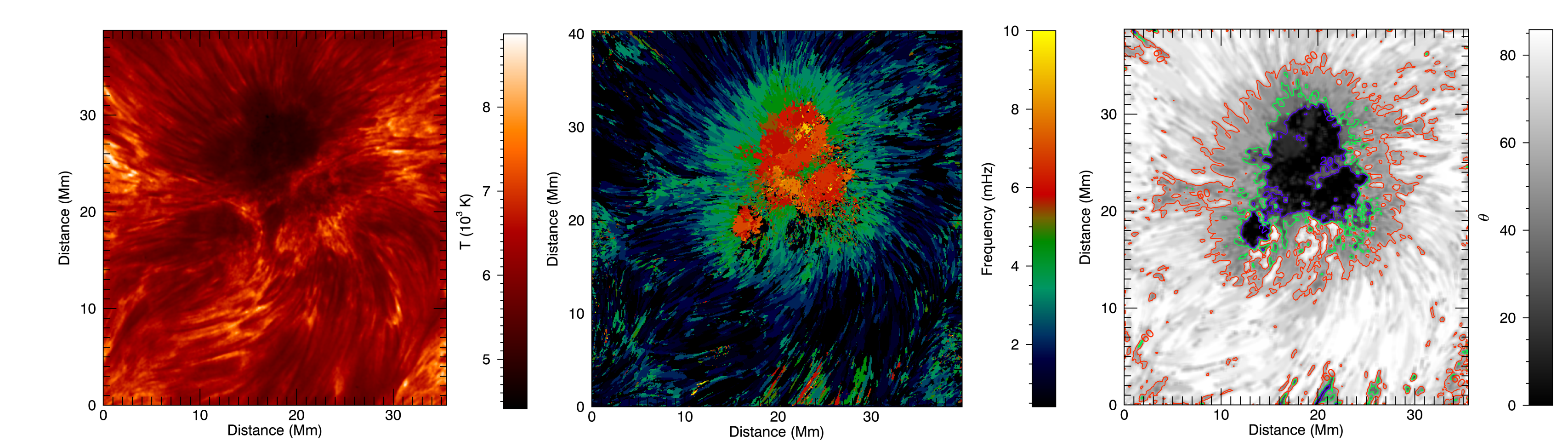}
\caption{Estimating magnetic field inclination in the sunspot. The left panel is the estimated temperature of the chromosphere. The middle panel shows the frequency values with the largest power in each pixel for the Doppler velocity time-series. The right-hand panel displays the estimated inclination angle of the magnetic field with respect to the vertical. The overplotted contours mark inclinations of 20$^\circ$ (blue), 40$^\circ$ (green) and 60$^\circ$ (red). Note that the inclination map has been averaged with $0''.7$ box-car filter to aid visualisation.}\label{fig:freq_dist}
\end{figure*}
%#############################################################################
%#############################################################################

\subsection{Slow wave magneto-seismology}

In advance of our discussion, we would like to obtain an impression for 
the inclination of the magnetic fields around the sunspot. If we take forward the assumption that the running 
penumbral waves are upwardly propagating slow waves along inclined field lines, \cite{BOGJUD2006} suggested 
that one can use the frequency values of the slow waves with the largest power, $\nu_{max}$, 
to map the inclination of the magnetic field \citep[see][for a full discussion and an implementation of this method]{LOHBOT2016}. 

\medskip
The inclination of the magnetic field, $\theta_B$ is given by
\begin{equation}
 \theta_B = \cos^{-1}\left( \frac{4\pi\nu_c}{g}\sqrt{\frac{R T}{\gamma\,\bar{\mu}}}\right)
 \label{eq:mag_inc}
\end{equation}
where $R$ is the molar gas constant, $\bar{\mu}$ mean molar mass (which we take as 1.26 for a weakly ionized solar plasma), $T$ is the plasma temperature. The key to using Equation~(\ref{eq:mag_inc}) is finding a relationship between $\nu_c$ and $\nu_{max}$. \cite{LOHBOT2016}
used an empirical relationship, $\nu_{max}\approx C_{emp}\nu_{c}$, where $C_{emp}=1.25$ was suggested by \cite{BOGJUD2006}. 
However, this relationship is based upon the maximum frequencies found at the umbral centre in the observations 
used by \cite{BOGJUD2006}, with the assumption that $\nu_c=5.2$~mHz and maximum frequencies are between 6-7~mHz. There 
will be some need to alter $C_{emp}$  based upon the observed values of maximum frequency found in each spot, 
otherwise the above formula does not perform well. The inversion relationship, ignoring the standard constants, 
comes down to 
\begin{equation}
 \cos\theta_B \propto \frac{\nu_{max}\sqrt{T}}{C_{emp}},
\end{equation}
and it is clear that an issue which can arise is 
relatively large values of $\nu_{max}$ lead to the RHS being larger than 1. The frequency values for which
this occurs will also be influenced by the estimated temperature too. This means that a modification of $C_{emph}$
is required to get the best results for the inversion. Here we find that $C_{emph}=1.35$ gives reasonable results for
this sunspot, although, umbral pixels with the largest $\nu_{max}$ ($>7$~mHz) still cause issues and inversion angles are set 
to 0 degrees. We discuss what we believe to be the reason for this in the following paragraph. 

\medskip

Applying Fourier analysis on a pixel-by-pixel basis across the FOV, maps are produced that show the frequency value at 
which the power of the oscillations is found to be largest (Figure~\ref{fig:freq_dist}). This is performed on both intensity and velocity data, however, it is 
found that the intensity power maps are dominated by the lowest frequencies outside of the umbra - potentially due to long-term variations in the sunspot atmosphere. Hence, unlike \cite{LOHBOT2016}, we use the maximum frequencies from the Doppler 
velocity data (Figure~\ref{fig:freq_dist} middle panel). In the umbral sections, higher-frequency magnetoacoustic 
power can be seen to be dominant, with most of the umbra having periodicities of $6-7$~mHz. As expected, the frequency of maximum 
power decreases in the transition from
umbra into penumbra. Away from the sunspot, low-frequency power dominates, which is probably due to 
long-term evolution of the plasma velocity. As mentioned above, a few umbral regions show the largest power is at frequencies greater than 7~mHz. Inspection of the data suggests that the
largest $\nu_{max}$ values come from locations where the Doppler velocity time-series has strong 
discontinuities, which could cause enhanced high-frequency power in a discrete Fourier transform. It would 
appear that these sharp changes in velocity are artefacts arising from the model fitting of the line 
profile. Given we are primarily interested in finding the inclination of the super-penumbral fibrils 
rather than the inclination across the spot, the pixels with affected Doppler time-series are not an 
issue. Needless to say, the interpretation of the results from these regions should be undertaken with
caution.

\medskip
\subsection{Temperature estimates}
{Temperature maps were produced using semi-empirical modelling with the Non-LTE Inversion COde using the Lorien Engine} 
(\verb|NICOLE|; \citealt{SOCASetal2015}).{Due to the large FOV and limited computing resources, only one inversion cycle was 
run. Per parameter, the following equidistant node scheme was used: Temperature - 8, LOS Velocity - 3, Microturbulence - 1, Magnetic 
field (each Cartesian component) - 1. The starting model was the ``FAL-C'' model \citep{Fontenla_1993} and the reference for line-core 
wavelength calibration was a patch of quiet Sun. The latter choice differs by only 120~m$\;$s$^{-1}$ from selecting a time average 
over the umbra for these observations \citep{2020arXiv200805482H}. NICOLE was run without modifications and the following key settings 
were selected: three rays for the non-LTE computation cycle with LTE starting populations (a compromise between speed and accuracy), 
cubic DELO-Bezier solver for the radiative transfer equation \citep{de_la_Cruz_Rodr_guez_2013}, and isotopes for the Ca atom model 
included as in \cite{Leenaarts_2014}. The inverted temperatures were averaged around the formation height of the Ca II 8542~{\AA} line 
core. For the penumbra, at a $\mu$ of 0.745, this height is around $\log\tau_{500}$=-5.5 \citep{Bose_2019}. Given the higher $\mu$ in 
this work we followed a simple average at $\log\tau_{500}$=-5$\pm$1. Very little noise is present in the temperature maps which is due 
to very few pixels showing strong local divergence of solutions (i.e. the ``inversion noise'' is low).}

Figure~\ref{fig:freq_dist} displays the estimated temperature (left panel) and LOS inclination estimated using Eq.~\ref{eq:mag_inc} 
(right panel). As can be seen in the LOS inclination map, 
the umbral field is near vertical and the angle to the vertical increases from the umbra to super-penumbra. The super-penumbral fibrils have inclinations around 40-70$^{\circ}$.

\medskip

 %#############################################################################
%#############################################################################
\begin{figure}[!tp]
\centering
\includegraphics[scale=0.62, clip=true, viewport=3.0cm 7.0cm 18.5cm 20.cm]{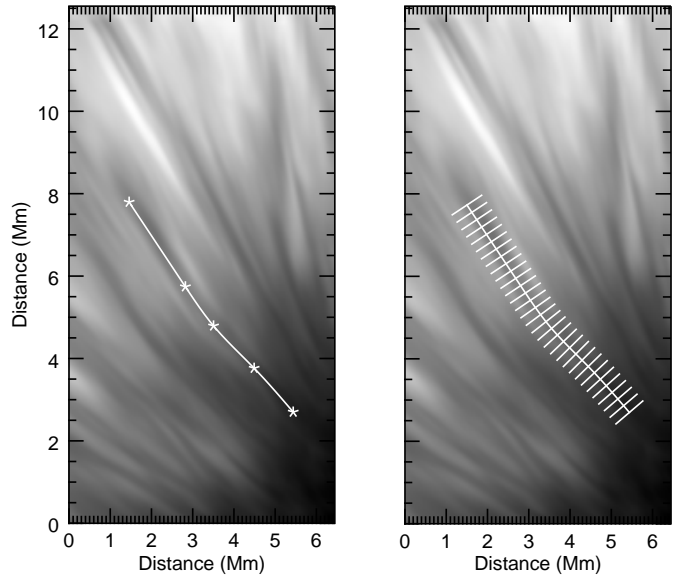}
\caption{An isolated set of fibrils in the Ca II sunspot super-penumbra. The left panel of the figure shows the manually selected guide points (white stars) and the resulting outline of the fibril's central axis is over-plotted. In the right panel, the cross-cuts normal to the guide-line is shown along the fibril structure. For clarity a small number of guide points and a large separation distance is chosen.
}\label{fig:fibril_slit}
\end{figure}
%#############################################################################
%#############################################################################

\section{Transverse motions}
Spanning out near-radially from the umbra are the chromospheric super-penumbral fibrils. Previous 
studies of the fibrils enables us to state some of their basic properties. Through 
spectro-polarimetric observations of 
Ca II 8542~{\AA}, \cite{ASEetal2017} demonstrated the dark fibrils in sunspot penumbra largely align with the magnetic field orientation. This is also supported by alignment 
between the thermal and magnetic structure inferred from He 10830~{\AA} observations of super-penumbral fibrils \citep{Schad_2013}. Moreover, \citet{Schad_2015} suggest that the 
lateral structure of the magnetic field throughout the penumbral chromosphere is relatively homogeneous, implying that the observed inhomogeneities are due to density or 
temperature. The reduced intensity of the fibrils suggest that the plasma has greater opacity at 8542~{\AA} (and in other chromospheric lines) than the surrounding plasma, 
suggesting it is denser than the ambient plasma.  Moreover, as was demonstrated with the slow mode magneto-seismology, the 
magnetic field aligned with the observed fibrils was found to 
be inclined from the vertical. The implication of this is that the fibrils are field-aligned 
density enhancements, providing wave-guides that can define MHD wave propagation.

\medskip

Due to high spatial resolution of this data set (and high signal-to-noise), we are able to follow the motions of these fibrils and find evidence of quasi-periodic transverse motions.

\subsection{Individual fibrils}
First, we focus on a few examples to demonstrate the features of the transverse motion. The examples are selected because {the fibril is visible for a significant portion of the data set.} 

In order to follow the 
propagation of the transverse motion, we trace out what we believe to be the magnetic axis. The time-average of the LPM images are used ({averaged over 2400~s}) to 
determine the average location of the fibrils and a curvilinear path is defined that follows the central axis. Figure~\ref{fig:fibril_slit} demonstrates an 
example of a cubic spline fitted to the longitudinal axis of the fibril. Cross-cuts normal to this curve are then 
taken, shown by the white lines in the right hand panel. 

%#############################################################################
%#############################################################################
\begin{figure}[!tp]
\centering
\includegraphics[scale=0.47, clip=true, viewport=1.cm 6.0cm 21.cm 21.5cm]{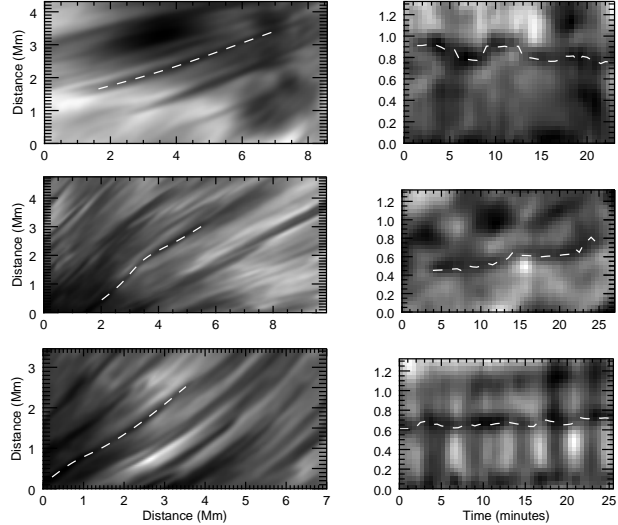}
\caption{Examples of a fibril showing transverse motions. The left-hand panel displays the fibril and the trace of its axis. The right hand panel is a time-distance diagrams from a perpendicular cross cut, with the dashed line indicating the central location of the fibril axis over time.}\label{fig:kink_three}
\end{figure}
%#############################################################################
%#############################################################################

\medskip
The cross-cuts are used to generate time-distance diagrams from the full LPM image time-series.  
Figure~\ref{fig:kink_three} displays time-distance diagrams generated from cross-cuts taken across three separate fibrils located around the sunspot. Each one is seen to support transverse motions~\footnote{We note that manifestations of the slow magnetoacoustic waves can be seen as intensity ridges (lower right panel Figure~\ref{fig:kink_three}). Features like this are most evident in time-distance plots that originate closer to the umbra.} (right panel), 
highlighted by the white curve that shows the central location of the fibril's axis. The location of the 
centre of the fibril axis is found by a fitting a Gaussian function to the {intensity values of the fibril's cross-sectional profile} in 
each time frame \citep[see e.g.,][]{MOR2014, MORetal2014,MOOetal2017}. {The uncertainties on the intensity values are used in the fitting process and estimated following the method described in \cite{MOOetal2017}. }

\begin{figure}[!tp]
\centering
\includegraphics[scale=0.55, clip=true, viewport=0.cm 0.0cm 17.cm 13.5cm]{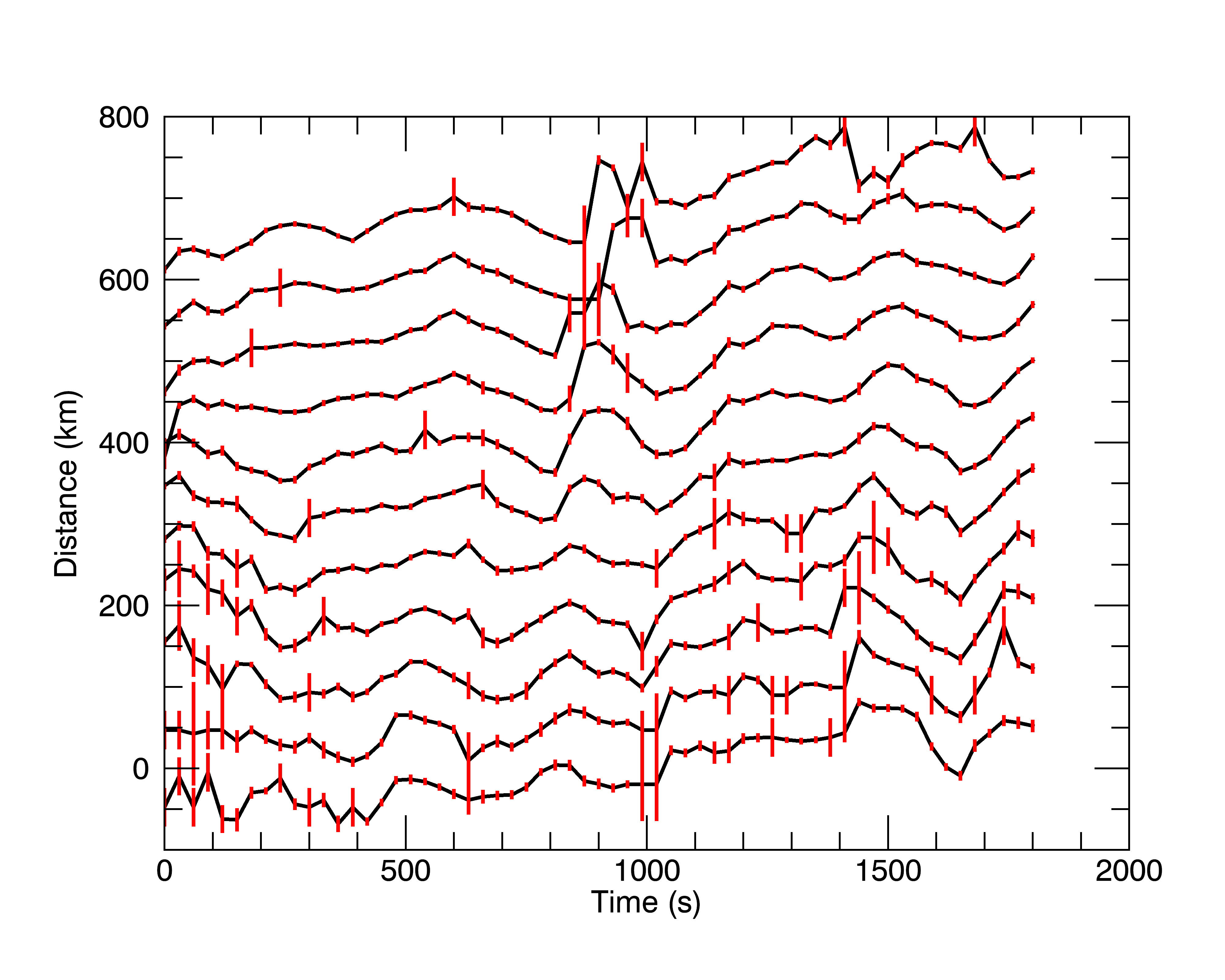}
\caption{Displacement time-series from along a single fibril shown in Figure~\ref{fig:fibril_slit}. Black curves relate to the measured central location of the fibril cross-section, while the red vertical lines show the 1 sigma uncertainties on each measurement. Each series is measured at 150~km intervals along the longitudinal axis of the fibril. The series are off-set by an arbitrary amount for visualisation}\label{fig:kink_series}
\end{figure}

\medskip
{The fibrils are observed to undergo a transverse displacement, with the displacements showing
coherent motion along the length of the fibril, i.e., the longitudinal axis moving nearly in unison. In
Figure~\ref{fig:kink_series} we show a series of displacement time-series for one fibril, each taken from
locations along a fibril's longitudinal axis.} As is seen in Figure~\ref{fig:kink_three} and Figure~\ref{fig:kink_series}, the displacements
of the fibril are typically relatively small, on the order of $\sim100$~km; which is similar to the 
scale of kink waves seen in chromospheric fibrils in other regions of the Sun's atmosphere \citep{MORetal2014,JESetal2015,JAFetal2016}. {The fibril is seen to support a number of short wave packets that lead to near-sinusoidal transverse motion\footnote{By near-sinusoidal we mean that the fibril is seen to sway back and forth, with a displacement time-series that could be predominantly described by a sinusoidal function. Although, typically there is some deviation from a pure sinusoid.}. The evolution of the wave packets can be seen to be relatively rapid, increasing or decreasing in amplitude over short distances. This behaviour has been seen previously in quiet Sun \citep{KURetal2013} and plage \citep{JAFetal2016} observations of chromospheric transverse motions. Given this, we interpret the motions observed here as signatures of transverse waves, namely the MHD kink mode.}

 %#############################################################################
%#############################################################################
%#############################################################################
%#############################################################################
 
\begin{figure}[!tp]
\centering
\includegraphics[scale=0.5, clip=true, viewport=2.cm 7.0cm 21.cm 20.cm]{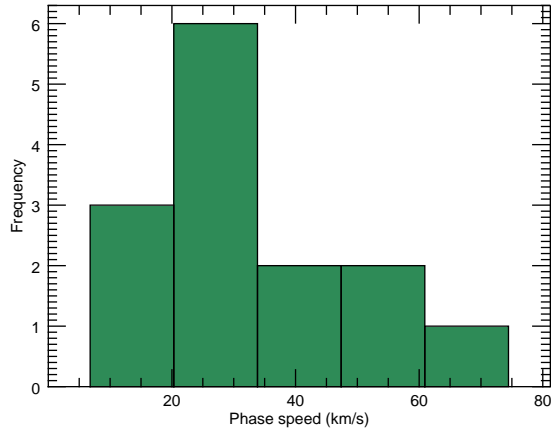}
\caption{Distribution of phase speeds. Histogram showing the distribution of phase speeds of the ten isolated super-penumbral fibrils.}\label{fig:kink_phase}
\end{figure}

\medskip
Following the method described in \cite{MOOetal2017}, we estimate propagation speeds for the {transverse motions} supported 
by ten individual fibrils. Interestingly, much like the internetwork fibrils examined in \cite{MOOetal2017}, 
the propagation speeds appear to be invariant over the fibrils' length. Figure~\ref{fig:kink_phase} 
displays a histogram of the measured propagation speeds and they 
are summarised in Table~\ref{tab:wave_prop}. The propagation speeds are less than that found in the quiet Sun and 
internetwork fibrils \citep{PIEetal2011, MORetal2012c, KURetal2013, MOOetal2017}, 
although they are similar to those found in in Ca II H slender fibrils \citep[SUNRISE -][]{JAFetal2016, 
GAFetal2016}. The expected sound speed in the sunspot chromosphere is $\sim6-10$~km\,s$^{-1}$ (estimated from the temperature inversions), and 
all measured values are larger than this. In principle, the larger magnetic field strengths in active regions \citep[$\sim300-400$~G -][]{ASEetal2017} should lead to larger Alfv\'en speeds than we observe. Given the inclination of the field lines, 
we are likely underestimating the length of the fibrils due to projection effects leading to smaller values of propagation speeds. 
Although, considering the map of inclination angles obtained for the sunspot 
(Figure~\ref{fig:freq_dist}), the length is likely underestimated by around 15-50~\% (inclination of 
40-70$^\circ$ to the vertical) - leading to a comparable increase in propagation speeds. 

\medskip

With the cadence of this data set at 30~s, the temporal resolution is really at the upper limit to provide 
a detailed analysis of the transverse motions. {In general, it is not possible with this data set to find many examples where the propagation of the transverse motions along the fibrils can be measured with certainty. Under the assumption of wave motion, e.g., the kink mode, the displacements would propagate at typical Alfv\'en speeds of $\sim100$'s~km~s$^{-1}$. Moving at such speeds would lead to small lags between neighbouring cross-cuts that would be difficult to find, at least with the current methodology.} Moreover, examining movies of the data reveal that the fibrils can disappear and reappear (this behaviour is also visible in the time-distance diagrams), much like 
the internetwork fibrils in \cite{MOOetal2017} but on longer time periods\footnote{We do not examine the reasons behind the fibril visibility, but suggest it could be the result of 
mass-loading events leading to variability in the fibril density \citep{CHAEetal2014,CHAEetal2015, LEEetal2015}, and hence opacity in Ca II.}. Hence this places further limits on
following wave motions. {This restriction is not limited to measuring propagation; for example, the dominant periods from measurements of transverse waves in other fibrils are around 100-150~s 
\citep{MORetal2014} corresponding to $\sim$3-5 time-frames here.}

\medskip

The transverse motions are not limited to the ten features studied in this section; they
are found to be ubiquitous through the super-penmbral fibrils. In order to provide a broader 
picture of the transverse motion in the super-penumbra, we will utilise an automated method to undertake further measurements \citep{WEBetal2018}.

\begin{table}
\centering
\caption{Super-penumbral fibril transverse wave properties \label{tab:wave_prop}}
\begin{tabular}{lccccc}
\hline\hline
  & Mean  & Median & Mode & $\sigma^\dagger$ & $n^*$\\ 
  \hline
Displacement (km) & 74  & 62 & 46 & 47 & 2306\\
Period (s) & 754  & 570 &  330 & 529 & 2306 \\
Velocity (km/s) & 0.76  & 0.66 & 0.34 & 0.47 & 2306\\
Prop. speed (km/s) & 25  & 21 & - & 0.74& 10 \\
\hline
\multicolumn{4}{l}{$^\dagger$ Standard deviation}\\
\multicolumn{4}{l}{$^*$ number of measurements}
\end{tabular}
\end{table}

 %#############################################################################
%#############################################################################
%#############################################################################
%#############################################################################

\subsection{General properties}

\begin{figure}[!th]
\centering
\includegraphics[scale=0.65, clip=true, viewport=0.0cm 0.0cm 24.0cm 10.5cm]{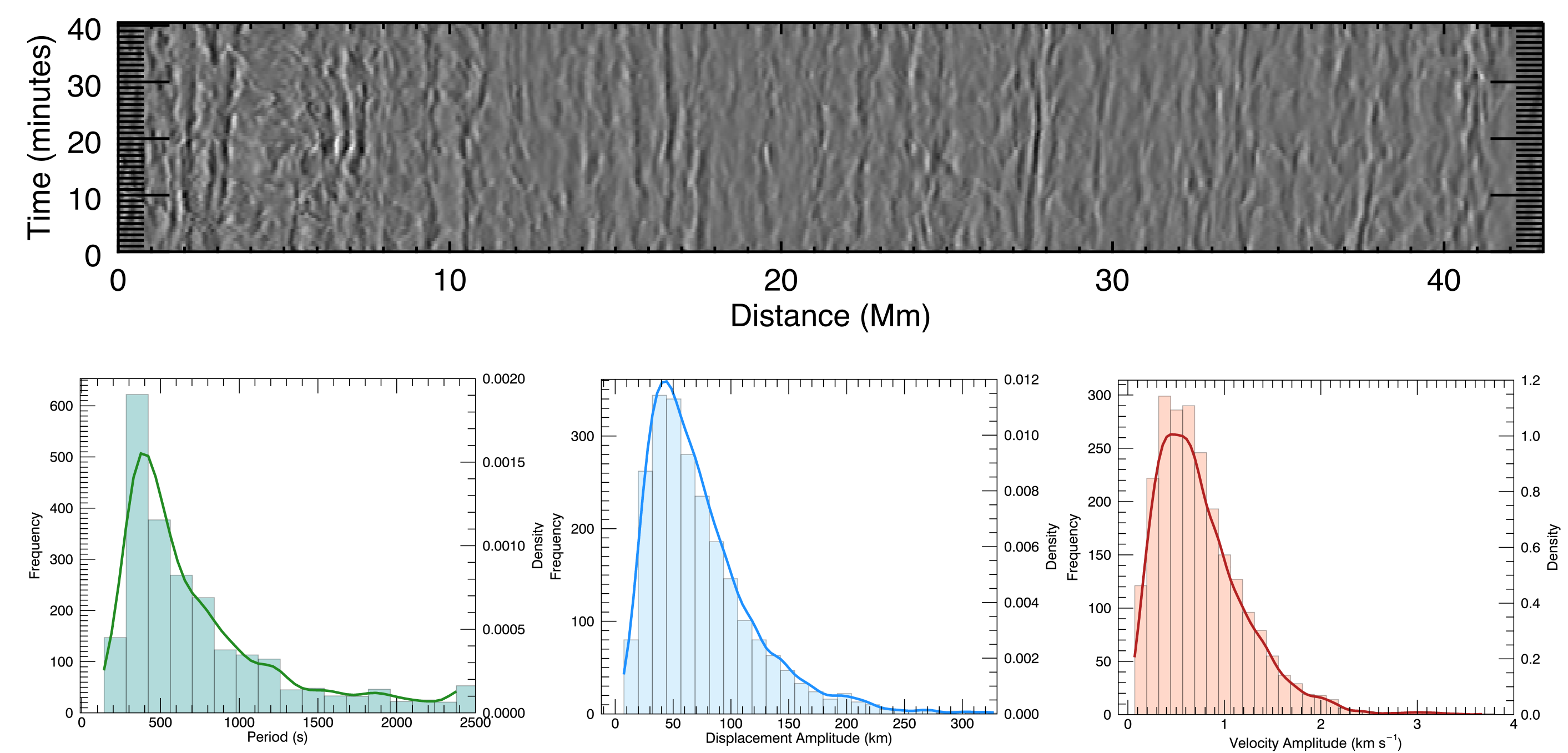}
\caption{Example time-distance diagram of super-penumbral fibrils. The transverse displacement of the fibrils is evident, 
with a number showing near sinusoidal displacements. The figure is generated from a arc-shaped cross-cut, with a radius of 12 Mm and centred on the sunspot bary-centre. The data has been unsharp masked. The bottom row displays 
summary plots of the transverse wave properties: period (left), displacement amplitude 
(middle) and velocity amplitude (right). Each panel displays the histogram, binned
using Scott's rule, and the kernel density estimate (solid line) with the bandwidth selected using
Silverman's rule. }\label{fig:kink_all}
\end{figure}

To reveal how wide-spread these {motions} are, we define arcs that are at a fixed distance from the sunspot umbral 
bary-centre. These arcs sample the fibrils located in upper half of the sunspot, where the fibrils are 
largely aligned radially. Hence, the tangents to the arc are largely perpendicular to the fibrils' longitudinal axis. The arcs do not cover the lower half of the spot as fibrils are more curved and 
not always radially orientated. To remove large-scale spatial variations of intensity and improve visibility of the fibrils, 
we implement two-dimensional square box-car filter of side-length $0''.9$ in order to unsharp mask the data. 
Figure~\ref{fig:kink_all} shows an 
example of the time-distance diagrams obtained from critically sampling the data along the arcs. 
By zooming in on the figure it can be seen that most of the dark super-penumbral fibrils are subject to 
some transverse motion, with some showing clear near-sinusoidal displacements.

To measure the typical properties of the kink waves, we use the NUWT code. We provide here a 
cursory summary of the methodology but refer readers to \cite{WEBetal2018} for a detailed 
discussion. In each time-distance diagram, the fibril centers are located. A sub-pixel estimate 
of the location is obtained by approximating the cross-sectional structure as a Gaussian and 
finding the best model parameters via weighted non-linear least squares. Uncertainties on 
intensity values were calculated via the methodology discussed in \cite{MOOetal2017}. By 
following each fibril in time, a displacement time-series can be constructed. Each of these 
time-series is then subject to a Fourier decomposition, and the amplitude and period of the 
dominant Fourier mode is obtained. A lower limit is placed on the length of 
time-series in order to keep noisy time-series to a minimum; the shortest time-series here 
are of length 10 samples.

%#############################################################################
%#############################################################################
%#############################################################################
%#############################################################################

Figure~\ref{fig:kink_all} (bottom panels) display summary plots of the basic properties of the transverse motions from the NUWT results, namely displacement and velocity amplitude,
and period. A statistical summary of the measures of centre and spread are also given in Table~\ref{tab:wave_prop}.
The typical displacements found here are broadly in agreement
with those found in other chromospheric fibrils \citep{MORetal2013, JESetal2015, JAFetal2016}, however the central 
measures for the periods and velocity amplitudes are 
larger and smaller, respectively, compared to the previously reported values. The dominance of longer periods here could be attributed to the relatively long cadence of the current data set (in comparison to previous
studies), meaning we cannot sample the parameter space effectively. Furthermore, given that velocity amplitude and period show a negative correlation\footnote{This trend has also been reported previously in studies of both chromospheric \citep{MORetal2013}, and coronal 
\citep{MORetal2019} transverse waves.}, the inability to properly measure the higher frequency waves would lead to a reduction in the central 
measures for velocity amplitude. This speculation would need to be confirmed with higher cadence data sets.

\medskip

\begin{figure}[!tp]
\centering
\includegraphics[scale=0.55, clip=true, viewport=0.5cm 0.0cm 16.5cm 12.5cm]{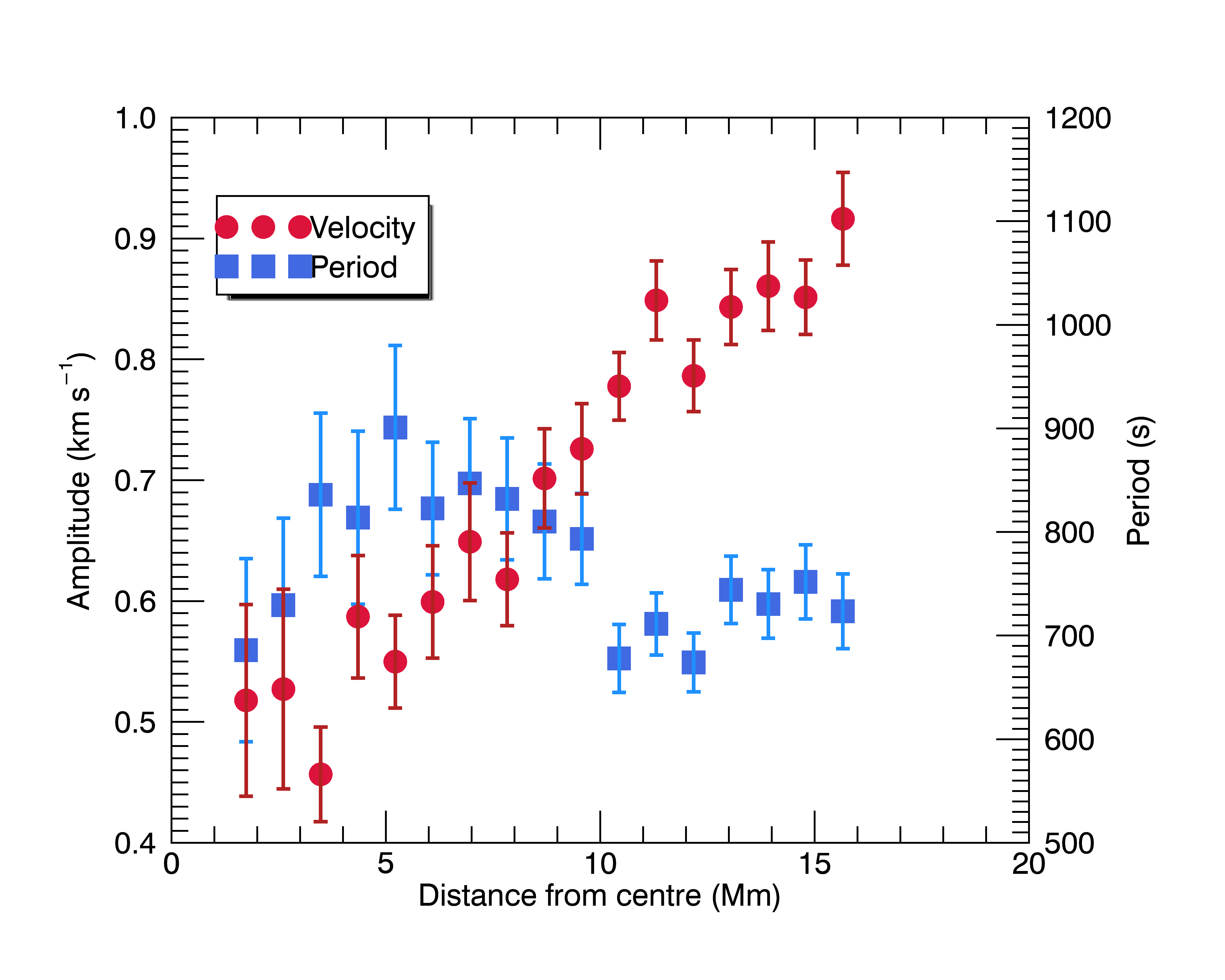}
\caption{Wave properties as a function of distance from the umbral centre. The figure displays the mean values for velocity and period, with the error bars denoting the standard errors on the mean.}\label{fig:kink_amp}
\end{figure}

Finally, we show an interesting result that requires further investigation but will be left to a future study. Figure~\ref{fig:kink_amp}
displays the mean velocity amplitude of the transverse waves as a function of distance from the umbral bary-centre. {The mean values are calculated from all measurements obtained in an individual arc.} The
measurements show that the mean value of the wave amplitude
increases with distance from the sunspot centre. Assuming that
the wave propagation can be suitably described by WKB theory,
then the variation in amplitude, $v$, can be associated with
a change in density, $\rho$. The velocity amplitude is 
expected to vary as, $v\propto \rho^{-1/4}$ \citep[assuming the waves are Alfv\'enic, see][]{MOR2014}. Hence, an increase in $v$ suggests that the density decreases along the 
fibrils. This is perhaps not unexpected, as the super-penumbral fibrils are likely tracing the magnetic field
as it rises out of the sunspot umbra and re-connecting with the
solar surface at some distance from the spot (apparently outside of the current SST field of view). It is naturally expected that the density would decrease with height. Given
the inverse relationship found between period and velocity amplitude, mentioned previously, we also show the mean period
in Figure~\ref{fig:kink_amp}. The mean value of period is consistent with a constant value as a function of radius, within 2 standard errors. Hence, the observed change in wave amplitude is not due to a change in the populations
of waves being measured.

%#############################################################################
%#############################################################################
%#############################################################################
%#############################################################################

\section{Discussion and Conclusions}

We have demonstrated a range of MHD wave phenomenon are present within the chromosphere of the sunspot under observation. However, our primary focus is on
the transverse motions of the super-penumbral fibrils, {which we have interpreted as signatures of transverse waves, namely the MHD kink mode}. 
This is not the the first observation of transverse wave modes in the chromosphere \citep[e.g.,][] {JESetal2015}, or even the first report of such waves in super-penumbral 
fibrils \citep{PIEetal2011}. Here we have demonstrated that, like seemingly everywhere else in the chromosphere, 
transverse waves modes pervade the sunspot chromosphere. We also characterised the properties of the super-penumbral transverse waves through a large-scale study. The presence of 
ubiquitous transverse waves in the sunspot atmosphere raises immediate questions about the nature of their excitation. The answer to these questions could provide important insights into the flow of energy through the lower atmosphere. We discuss the potential candidates in the following.

\subsection{Convection-driven}
Conventionally, transverse modes are expected to be driven by convective motions buffeting magnetic field concentrations in the photosphere, with evidence for this relationship between photospheric granular motions and chromospheric waves in the quiet Sun \citep{MORetal2013}.
However, the strong magnetic field that forms within the sunspot acts to suppress the convective motions, with the suppression especially evident in the umbra. 
The strong field is also thought to modify convection at the periphery of the spot, forming the penumbra and leading to a highly directional convection. Moreover, the measured properties of the transverse waves here reveal typically larger periods and smaller velocity amplitudes than their quiet Sun counterparts.
If one where to assume that convection drove the transverse waves with the same energy as in the quiet Sun, the differences in velocity amplitude could be attributed to 
stronger magnetic fields in the sunspot. On the other hand, the difference in typical periods could indicate a 
non-convective excitation. As mentioned, we note that we do 
not believe that we are able to effectively measure high 
frequency waves; hence the measured distribution 
(Figure~\ref{fig:kink_all}) might not reflect the true distribution of 
periods. Given the modified convection within the sunspot, 
this raises the question as to whether another mechanism might be dominant in exciting the transverse modes found 
along the super-penumbral fibrils?

\subsection{Mode-conversion}
As discussed in the introduction, mode conversion is excepted to be a natural feature in the sunspot. It is now well established (at least numerically) that $p$-modes can convert to either slow or fast magnetoacoustic waves at the equipartition layer, depending upon the local 
inclination of the magnetic field
\citep[e.g.,][]{BOGetal2003, Khomenko_2006, SCHCAL2006}. The model of \citet{Khomenko_2012} suggests that the generation of upward propagating fast
modes is most efficient for inclinations of 30$^\circ$ - 60$^\circ$ (c.f. their Figure~5). Such values are consistent with the inclination of the super-penumbral magnetic field estimated through the slow wave magneto-seismology (Figure~\ref{fig:freq_dist}). Additionally, it is expected that in sunspot atmospheres, the fast modes generated at the
equipartition layer are reflected due to gradients in the 
Alfv\'en speed near the transition region. Although it is worth noting that, to date, the models that examine wave propagation in sunspots do not incorporate the inhomogenous 
plasma structures observed in the chromosphere, i.e., the super-penumbral fibrils. The lack of small-scale waveguides
in the simulations means that the fast mode propagation is not restricted to follow magnetic field lines.
This is not the case for structured media, where fast mode 
wave energy can be `trapped' within density enhancements and 
hence will follow the magnetic field \citep[e.g.,][]{VANetal2008b}. This could imply that a fraction of the fast
mode wave energy are not reflected along the 
inclined field lines, as they do not reach the transition 
region; instead are siphoned along the chromospheric fine 
structure.

Furthermore, a number of recent theoretical and numerical investigations have shown that, the $p$-mode-excited fast magnetoacoustic modes can convert to Alfv\'enic modes at regions where sharp gradients in Alfv\'en speed exist, e.g.,
near the transition region \citep{CAL2011, Khomenko_2012}. There are indications that this 
latter mode conversion is responsible for part of the coronal flux of Aflv\'enic waves \citep{CAL2017, MORetal2019}. Whether this would be relevant for wave excitation along the superpenumbral fibrils is still unclear.

The transverse motions that we observe can be described by the kink mode. The kink mode can have qualities that means it is often described as Alfv\'enic in character \citep[e.g.,][]{GOOetal2012}, although it is still considered to be a fast 
magnetoacoustic mode \citep[e.g.,][]{DEMNAK2012}. Given the 
hybrid nature of the mode, it is possible that it could be excited at either the equipartition layer or at the transition region. However, only the former situation would seemingly agree with our observation. The analysis suggests that the waves are propagating in the chromosphere and away from the umbra, hence seemingly excited from below.

It is also worth providing a crude estimate of the wave energy in order to evaluate whether mode-conversion from $p$-modes is a valid candidate. The time average energy in the transverse modes can
be approximated as:
\begin{equation}
    E=\frac{1}{2}\rho v^2 v_A.
\end{equation}
We use the Alfv\'en speed, $v_A$, as opposed to the kink speed, in order to provide straightforward estimates for the wave propagation speed from values of $B$ and $\rho$ obtained through various methods. Given the uncertainties with the following estimate, the difference between using the kink speed and $v_A$ is negligible.
The magnetic field strength in the super-penumbral fibrils is on the order of 300-400~G \citep{ASEetal2017} and density estimates for chromospheric plasma at a height of 
$\sim1300$~km is $\sim1.3\times10^{-8}$~kg\,m$^{-3}$ \citep{MALTBYetal1986}. Using these values gives an Alfv\'en speed
between 200-300~km\,s$^{-1}$. We note this is around 10 times the propagation speed we are able to measure (Table~\ref{tab:wave_prop}). The reason for this significant difference is a mystery to us. As mentioned, \citet{JAFetal2016} also report small propagation speeds in fibrils in an active region. 

Given the disparity in expected and measured speeds, we have a broad range for $v_A$ possible (20-300~km\,s$^{-1}$). Energy fluxes are calculated to be somewhere between $E=80-1200$~W\,m$^{-2}$, using the mean value of velocity amplitude (Table~\ref{tab:wave_prop}). These values are broadly consistent with the amount of energy found in fast-modes through mode conversion from the simulations of \citet{KHOCAL2012}, $\sim300$~W\,m$^{-2}$.  
\medskip

\subsection{Reconnection driven}
A final alternative is wave excitation via magnetic reconnection. \citet{PIEetal2011} analyse a single transverse wave event that  demonstrates a relatively large amplitude 
(displacement $\sim 150$~km). Based upon other indicators in
their data set, the authors suggest that the wave was driven by a reconnection event. Moreover, various studies have demonstrated that there are potentially a significant number of reconnection events in the penumbra that lead to the presence of `jets' in the penumbral chromosphere \citep{KATetal2007}. These so called penumbral microjets are typically short in length and the associated enhancement of emission in Ca II lines is short-lived \citep[$1-2$~minutes - ][]{Vissers_2015,Drews_2017}. However, evidence seems to suggest there is 
little mass-motion associated with these features and they are potentially heating fronts associated with reconnection occurring lower in the atmosphere \citep{Esteban_Pozuelo_2019,Rouppe_van_der_Voort_2019,2020arXiv200502608D}. At present, it is unclear whether there is a connection between the occurrence of these microjets and the transverse waves - although the current data set may enable such a comparison to take place.

\subsection{Conclusions}
The results presented here confirm the ubiquity of transverse
wave motions throughout the chromosphere, adding sunspot super-penumbral fibrils to the increasing list of structures found to host the waves. However, their very presence in the
super-penumbral fibrils raises questions about their excitation that might not be asked about the transverse waves
found in other chromospheric features. Considering the options, we suggest mode conversion is the underlying mechanism that leads to the transverse waves in sunspots. We expect further investigation with higher cadence data will help shed some light on this.

% \aucontribute{RJM and KM performed analysis of the data. RJM drafted the manuscript. VMJH designed the observations, took the data
% and performed the data reduction. All authors read and approved the manuscript.}

% \competing{The authors declare that they have no competing interests.}

\acknowledgements{R.J.M acknowledges UKRI for support under the Future Leaders Fellowship MR/T019891/1. V.M.J.H would like to acknowledge support through the Research Council of Norway, project number 250810, and through its Centers of Excellence scheme, project number 262622. He also benefited from funding from the European Research Council (ERC) under the European Union’s Horizon 2020 research and innovation programme (grant agreement No. 682462). {We are deeply indebted to H. Socas-Navarro for continuously making his inversion code available to the community.}}

\end{document}